\documentclass[reprint,superscriptaddress,aps,amsmath,amssymb,showpacs]{revtex4-1}

\usepackage{graphicx}
\usepackage{bm}
\usepackage{cases}
\usepackage{amsmath}
\usepackage{color}
\usepackage{soul}

\newcommand{\ket}[1]{| #1\rangle}
\newcommand{\bra}[1]{\langle #1|}

\newcommand{\dket}[1]{| #1\rangle\!\rangle}
\newcommand{\dbra}[1]{\langle\!\langle #1|}
\newcommand{\dbraket}[2]{\langle\!\langle #1|#2\rangle\!\rangle}


\begin{document}
\title{Nonadiabaticity in spin pumping under relaxation}

\author{Kazunari Hashimoto}
\email[Electric adress: ]{hashimotok@yamanashi.ac.jp}
\affiliation{
Graduate School of Interdisciplinary Research, University of Yamanashi, Kofu 400-8511, Japan
}
\author{Gen Tatara}
\affiliation{
RIKEN Center for Emerging Matter Science (CEMS), 2-1 Hirosawa, Wako 351-0198, Japan
}
\author{Chikako Uchiyama}
\affiliation{
Graduate School of Interdisciplinary Research, University of Yamanashi, Kofu 400-8511, Japan
}

\date{\today}

\begin{abstract}
Using a minimum model consisting of a magnetic quantum dot and an electron lead, we investigate spin pumping by its precessing magnetization.
Focusing on the ``adiabaticity'', which is quantified using a comparison between the frequency of precession and the relaxation rate of the relevant system, we investigate the role of nonadiabaticity in spin pumping by obtaining the dependence of the spin current generated on the frequency of precession using full counting statistics.
This evaluation shows that the steady-state population of the quantum dot remains unchanged by the precession owing to the rotational symmetry about the axis of precession.
This implies that in the adiabatic limit the spin current is absent and that spin pumping is entirely a nonadiabatic effect.
We also find that the nonadiabatic spin current depends linearly on the frequency in the low-frequency regime and exhibits an oscillation in the high-frequency regime.
The oscillation points to an enhancement of spin pumping by tuning the frequency of precession.
\end{abstract}


\maketitle

\section{Introduction}

The generation of a spin-polarized electron current (spin current) in a normal metal without a bias voltage is one of the central issues in realizing spintronics applications \cite{wolf,maekawa}.
Standard models used in analyzing spin current generation, or so-called spin pumping, consists of two essential features: a) attaching a system such as a ferromagnet and/or a quantum dot to a normal metal lead and b) periodically modulating parameter(s) of the composite system.
Conventionally, the models proposed in analyzing spin pumping fall roughly into three classes: those using
(i) the precession of the magnetization in a ferromagnet attached to a normal metal lead \cite{tserkovnyak1,tserkovnyak2,wang,pzhang,fransson,winkler,rojek1,jahn,chen,tatara},
(ii) the periodic modulation of parameters such as gate voltages and tunneling amplitudes in a system that consists of quantum dots attached to normal metal leads \cite{mucciolo,cota,rojek2,riwar,braun,nakajima}, and 
(iii) the periodic modulation of the strength of magnetization in addition to parameters in a system consisting of a quantum dot attached to a ferromagnetic lead and a normal metal lead \cite{winkler,rojek1,splettstoesser}.
Of these models, those using the precession of the magnetization, model (i), have attracted intensive study as they can generate pure spin currents without any associated charge current.
This is because of the conservation of charge in the lead and a rotational symmetry about the axis of precession.
In this paper, we focus on the advantages and benefits of model (i).
We briefly summarize these conventional studies.

Spin pumping has been mostly studied in situations where the precession of the magnetization is sufficiently slow. Referred as adiabatic spin pumping, it was first proposal by Tserkovnyak et al. \cite{tserkovnyak1,tserkovnyak2} based on the scattering theory for adiabatic pumping given by Brouwer \cite{brouwer}.
Later, Wang et al. \cite{wang} derived an expression for the spin current with a finite precession frequency using Green's function techniques, and they showed that the expression reduces to its adiabatic expression, which in the low-frequency limit depends linearly on the frequency.
Recently, an expression for the spin current was derived from linear response theory \cite{tatara,chen}.
In these conventional investigations, the condition defining ``adiabaticity'' assumes that the modulation frequency $\Omega$ associated with the varying of the parameter (precession for spin pumping) is sufficiently small compared with the characteristic energy scale $\delta E$ over which the stationary scattering matrix or Green's function changes significantly \cite{moskalets,moskalets_text}; i.e.,
	\begin{equation}
	\Omega\ll \frac{\delta E}{\hbar}.
	\label{eq:condition1}
	\end{equation}
However, there is an another widely used condition for adiabaticity written
	\begin{equation}
	\Omega\ll\tau_{R}^{-1},
	\label{eq:condition2}
	\end{equation}
which compares the modulation frequency with the relaxation time of the relevant system $\tau_{R}$ \cite{nakajima,sinitsyn,hanggi,yuge,uchiyama}.
Physically, as for quasistatic processes in thermodynamics, the modulation frequency needs to be sufficiently slow enabling the steady state of the relevant system to follow it.
If the steady-state population of the relevant system explicitly depends on the modulating parameter, a certain fraction of the observable quantity is transferred through the system as a response to the change in population of the relevant system by the parameter modulation.
By summing up the transferred quantity over one modulation cycle, one may obtain a net amount of pumped quantity under the condition (\ref{eq:condition2}).
Adiabatic pumping under condition (\ref{eq:condition2}) was originally introduced by Sinitsyn and Nemenman in Ref. \cite{sinitsyn}, which presents a study of a stochastic kinetic equation describing a chemically reactive system.
It has been extended to quantum systems \cite{hanggi} and is used to study adiabatic pumping of an electron or energy system that is subjected to the modulation of environmental parameters \cite{nakajima,yuge,uchiyama}.
In Ref. \cite{nakajima}, it has also been used to study adiabatic spin pumping under amplitude modulation of a colinear magnetic field applied to both a quantum dot and leads.
To the best of our knowledge, spin pumping by precession of the magnetization has not yet been explicitly studied from the viewpoint of the adiabaticity condition in regard to the relaxation time, Eq. (\ref{eq:condition2}).
Because setting the relaxation time to infinity is impossible, it may be necessary to evaluate the effect of its finiteness on spin pumping, as well.

The main purpose here is to investigate spin pumping in situations where the relaxation time is comparable with or shorter than the period of precession, i.e.,
	\begin{equation}
	\Omega\lesssim\tau_{R}^{-1}.
	\end{equation}
For this purpose, we consider a minimum model \cite{tatara} consisting of a magnetic quantum dot attached to an electron lead and analyze spin pumping by its precessing magnetization.
Using full counting statistics \cite{levitov,utsumi,esposito} with the quantum master equation \cite{nakajima,yuge,uchiyama}, we obtain a frequency dependence for the spin current and an underlying electron dynamics.
In the formulation, we take into account all contributions including empty and completely filled states of the dot in the dynamics as well as spin flips between half-filled states of the dot; in conventional studies only the latter is considered \cite{wang,tatara}.
We find that all contributions are important for spin pumping.
With this formulation, we can cover the range from the low-frequency limit up to the order of the relaxation rate $\tau_{R}^{-1}$.
Surprisingly, we find that the steady-state population of the quantum dot remains unchanged under precession in contradistinction to that encountered in conventional studies on the adiabatic limit, Eq. (\ref{eq:condition2}) \cite{nakajima,sinitsyn,hanggi,yuge,uchiyama}.
This arises from the rotational symmetry about the axis of precession of the model (i).
We also find that population conservation leads to the absence of a spin current in the adiabatic limit under condition (\ref{eq:condition2}).
This implies that spin pumping is entirely a nonadiabatic effect when viewed from this perspective.
Also we find that the nonadiabatic spin current linearly depends on the frequency in the low-frequency regime, and exhibits oscillations in the high-frequency regime.

The paper is organized as follows: we introduce our minimum model (Sec. II) and summarize the full counting statistics (Sec. III.A), using it to formulate spin pumping (Sec, III.B). We consider spin pumping in the adiabatic limit (Sec. IV), and a numerically study of it (Sec. V) without using the adiabatic approximation. In Sec. VI, we provide discussions and concluding remarks.

\section{Model}
With our spin pumping model of a ferromagnetic quantum dot attached to an electron lead \cite{tatara,wang}, the dot has a dynamic magnetization ${\bf M}(t)$ that rotates around a fixed axis (labeled $z$-axis).
An electron in the dot is spin polarized because of the {\it s--d} exchange interaction with the magnetization, and is represented by two-component creation and annihilation operators ${\bf d}^{\dagger}=(d^{\dagger}_{\uparrow}$, $d^{\dagger}_{\downarrow})$ and ${\bf d}$, where $\uparrow$ or $\downarrow$ represents the spin polarization of the electron parallel or anti-parallel to the $z$-axis.

The Hamiltonian of the model has three term $H(t)=H_{{\rm d}}(t)+H_{{\rm l}}+H_{{\rm t}}$.
Here the term $H_{\rm{d}}(t)$ governing the dot is defined by
	\begin{equation}
	H_{{\rm d}}(t)={\bf d}^{\dagger}(\epsilon_{{\rm d}}-{\bf M}(t)\cdot{\boldsymbol\sigma}){\bf d},
	\end{equation}
where ${\bf M}(t)\equiv M(\sin\theta(t)\cos\phi(t),\sin\theta(t)\sin\phi(t),\cos\theta(t))$ and $\epsilon_{{\rm d}}$ is the unpolarized energy of a quantum-dot electron.
Introducing the eigenstates $\ket{j_{\uparrow},j_{\downarrow}}$ (with $j_{\sigma}=0$ or $1$) of the number operator of the dot $\sum_{\sigma}d^{\dagger}_{\sigma}d_{\sigma}$ as a basis, the dot Hamiltonian $H_{\rm{d}}(t)$ is represented by matrix
	\begin{widetext}
	\begin{equation}
	H_{\rm{d}}(t)=\bordermatrix{
& \ket{0,0} & \ket{0,1} & \ket{1,0} & \ket{1,1} \cr
& 0 & 0 & 0 & 0 \cr
& 0 & \epsilon_{\rm{d}}+M\cos\theta(t) & -Me^{+i\phi(t)}\sin\theta(t) & 0 \cr
& 0 & -Me^{-i\phi(t)}\sin\theta(t) & \epsilon_{\rm{d}}-M\cos\theta(t) & 0 \cr
& 0 & 0 & 0 & 2\epsilon_{\rm{d}} \cr
           }.
    \label{eq:matrix-dotH}
	\end{equation}
	\end{widetext}
The electron lead is described by the term
	\begin{equation}
	H_{{\rm l}}\equiv\sum_{\sigma=\uparrow,\downarrow}\sum_{k}
\epsilon_{k}c^{\dagger}_{\sigma,k}c_{\sigma,k},
	\end{equation}
where $c_{\sigma,k}^{\dagger}$ and $c_{\sigma,k}$ ($\sigma=\uparrow$ or $\downarrow$) are the creation and annihilation operators of the lead electrons which are treated as free electrons with energy $\epsilon_{k}$.
The coupling between the dot and the lead is assumed to be spin conserving,
	\begin{equation}
	H_{{\rm t}}\equiv\sum_{\sigma=\uparrow,\downarrow}\sum_{k}\hbar v_{k}
(d^{\dagger}_{\sigma}c_{\sigma,k}+c_{\sigma,k}^{\dagger}d_{\sigma}),
	\end{equation}
where $\hbar v_{k}$ is the dot--lead coupling strength, which we assume to be weak.

\section{Formulation}

We define the spin current generated by a cyclic precession of the magnetization using full counting statistics \cite{esposito}, a brief summary of which is provided followed by its application in formulating spin pumping.

\subsection{Full counting statistics}

Consider a Hamiltonian $H=H_{0}+H_{{\rm int}}$ with $H_{0}\equiv H_{S}+H_{E}$ describing a general interacting system composed of a relevant system $S$ and an environment $ E$, and $H_{{\rm int}}$ describing their interaction.
The full counting statistics provides the time evolution of the transfer of a quantity from the relevant system to the environment using the difference between the outcomes of the two point projection measurement of an observable of the environment $Q$.
Denoting the measurement outcomes of $Q$ at $t_{i}$ and $t>t_{i}$ as $q_{t_{i}}$ and $q_{t}$, the net amount of the quantity transferred is given by difference $\Delta q\equiv q_{t}-q_{t_{i}}$, where its sign is chosen to be positive when the quantity is transferred from $S$ to $E$.
The statistics of $\Delta q$ is summarized in its probability distribution function
	\begin{equation}
	P_{t}(\Delta q)\equiv\sum_{q_{t},q_{t_{i}}}
\delta(\Delta q-(q_{t}-q_{t_{i}}))P[q_{t},q_{t_{i}}],
	\label{eq:PDF}
	\end{equation}
with joint probability to obtain outcomes $q_{t_{i}}$ and $q_{t}$, successively,
	\begin{equation}
	P[q_{t},q_{t_{i}}]\equiv{\rm Tr}[P_{q_{t}}U(t,t_{i})P_{q_{t_{i}}}
W(t_{i})P_{q_{t_{i}}}U^{\dagger}(t,t_{i})P_{q_{t}}],
	\label{eq:JP}
	\end{equation}
where ${\rm Tr}$ is the trace taken over the total system, $P_{q_{t}}=|q_{t}\rangle\!\langle q_{t}|$ signifies the projective measurement of $Q$ at $t$, $U(t,t_{i})$ is the evolution operator of the total system, and $W(t_{i})$ is the initial condition of the total system.
The cumulants of $\Delta q$ are provided by its cumulant generating function
\begin{equation}
	S_{t}(\lambda)=\ln\int P_{t}(\Delta q)e^{i\lambda\Delta q}d\Delta q,
	\end{equation}
where $\lambda$ is the counting field associated with $Q$; e.g., the first cumulant, the mean, is computed from
\begin{equation}
	\langle\Delta q\rangle_{t}
=\frac{\partial S_{t}(\lambda)}{\partial(i\lambda)}\biggr|_{\lambda=0}.
	\label{eq:firstcum}
	\end{equation}

The full counting statistics provides a systematic procedure to obtain $S_{t}(\lambda)$.
Using the definitions (\ref{eq:PDF}) and (\ref{eq:JP}), and introducing the modified evolution operator $U_{\lambda}(t,t_{i})\equiv e^{i\lambda Q}U(t,t_{i})e^{-i\lambda Q}$ as well as the notation ${\bar W}(t_{i})\equiv\sum_{q_{t_{i}}}P_{q_{t_{i}}}W(t_{i})P_{q_{t_{i}}}$, $S_{t}(\lambda)$ is expressed as
	\begin{equation}
	S_{t}(\lambda)=\ln{\rm Tr}_{S}[\rho^{(\lambda)}(t)],
	\label{eq:cgf}
	\end{equation}
with
	\begin{equation}
	\rho^{(\lambda)}(t)\equiv
{\rm Tr}_{E}[U_{\lambda/2}(t,t_{i}){\bar W}(t_{i})U_{-\lambda/2}(t,t_{i})],
	\end{equation}
where ${\rm Tr}_{S}$ and ${\rm Tr}_{E}$ are the partial traces taken over the system and the environment, respectively.
Note that for $\lambda=0$, $\rho^{(\lambda)}$ reduces to the reduced density matrix of the quantum dot as $\rho^{(0)}={\rm Tr}_{E}[W(t)]$.
When the initial condition of the total system is a factorized state $W(t_{i})=\rho(t_{i})\otimes\rho^{{\rm eq}}_{E}$, where $\rho^{{\rm eq}}_{E}$ is the Gibbs state of the environment, the time evolution of $\rho^{(\lambda)}(t)$ is described by
	\begin{equation}
	\frac{\partial}{\partial t}\rho^{(\lambda)}(t)=\xi^{(\lambda)}(t)\rho^{(\lambda)}(t),
	\label{eq:QME}
	\end{equation}
which is the time-convolutionless-type quantum master equation \cite{kubo,shibataetal,hashitsume,uchiyama99} that has been modified to include the counting field \cite{uchiyama}.
Here, $\xi^{(\lambda)}(t)$ is a superoperator that acts on the density matrix $\rho^{(\lambda)}(t)$ and generates its time evolution.
With the Markovian approximation taken to second order in the interaction $H_{\rm{int}}$ and the time dependence of the generator omitted, its explicit form is then given by
	\begin{eqnarray}
	\begin{split}
	\xi^{(\lambda)}\rho&\equiv\frac{1}{i\hbar}[H_{S},\rho]\\
	&\;-\frac{1}{\hbar^{2}}\int^{\infty}_{0}d\tau{\rm Tr}_{E}[H_{{\rm int}},
[H_{{\rm int}}(-\tau),\rho\otimes\rho^{{\rm eq}}_{E}]_{\lambda}]_{\lambda},
	\label{eq:generator}
	\end{split}
	\end{eqnarray}
where $H_{{\rm int}}(t)\equiv e^{iH_{0}t/\hbar}H_{{\rm int}}e^{-iH_{0}t/\hbar}$ and $[A,B]_{\lambda}\equiv A^{(\lambda)}B-BA^{(-\lambda)}$ with $A^{(\lambda)}\equiv e^{i\lambda Q/2}Ae^{-i\lambda Q/2}$.

To work with the superoperator, it is convenient to introduce its supermatrix representation, where we represent the density matrix $\rho^{(\lambda)}$ in vector form and the superoperator $\xi^{(\lambda)}$ in matrix form (see Appendix \ref{app:supermatrix}).
The formal solution of the master equation (\ref{eq:QME}) is expressed as
	\begin{equation}
	\dket{\rho^{(\lambda)}(t)}=\exp\Bigr[\Xi^{(\lambda)}(t-t_{i})
\Bigr]\dket{\rho^{(\lambda)}(t_{i})},
	\label{eq:solution}
	\end{equation}
where $\dket{\rho^{(\lambda)}(t)}$ is the vector form of $\rho^{(\lambda)}(t)$ and $\Xi^{(\lambda)}$ is a supermatrix form of $\xi^{(\lambda)}$.
The cumulant generating function Eq. (\ref{eq:cgf}) is rewritten as $S_{t}(\lambda)=\ln\dbraket{1}{\rho^{(\lambda)}(t)}$, where $\dbra{1}$ is the trace operation acting to the right as $\dbraket{1}{\rho^{(\lambda)}(t)}={\rm Tr}_{{\rm d}}[\rho^{(0)}(t)]$.
Hence, the first cumulant of $\Delta q$ given by Eq. (\ref{eq:firstcum}) is rewritten as
	\begin{equation}
	\langle\Delta q\rangle_{t}=\biggr\langle\!\!\!\biggr\langle1\biggr|\biggr[
\frac{\partial}{\partial(i\lambda)}\rho^{(\lambda)}(t)\biggr]_{\lambda=0}
\biggr\rangle\!\!\!\biggr\rangle,
	\label{eq:1cumulant}
	\end{equation}
where we have used the invariance of the trace $\dbraket{1}{\rho^{(0)}(t)}={\rm Tr}_{S}[\rho^{(0)}(t)]=1$.
As $\dbraket{1}{\rho^{(0)}(t)}=1$, the state $\dbra{1}$ is a left-eigenstate of $\Xi^{(0)}$ with zero eigenvalue, i.e., $\dbra{1}\Xi^{(0)}=0$.
Using it together with Eqs. (\ref{eq:solution})--(\ref{eq:1cumulant}), we derive a formula for the mean of the transferred $Q$ during the time interval $t-t_{i}$ \cite{uchiyama},
	\begin{equation}
	\langle\Delta q\rangle_{t}=\int^{t}_{t_{i}}dt'\dbra{1}\biggr[
\frac{\partial\Xi^{(\lambda)}}{\partial(i\lambda)}\biggr]_{\lambda=0}
\dket{\rho^{(0)}(t')}.
	\label{eq:formula}
	\end{equation}
Its time derivative
	\begin{equation}
	J(t)\equiv\frac{d\langle\Delta q\rangle_{t}}{dt}=\dbra{1}\biggr[
\frac{\partial\Xi^{(\lambda)}}{\partial(i\lambda)}\biggr]_{\lambda=0}
\dket{\rho^{(0)}(t)}
	\end{equation}
provides the flow of the quantity $Q$ between the relevant system and the environment. Conversely, the integration of $J(t)$ over a given time interval provides $\langle\Delta q\rangle_{t}$.

\subsection{Spin current}

Based on Eq. (\ref{eq:formula}), we now formulate the spin current generated by the cyclic precession of the magnetization with period ${\cal T}$.
For the purpose, we consider the number of electrons with spin $\sigma(=\uparrow$ or $\downarrow)$ in the lead, represented by $N_{\sigma}=\sum_{k}c^{\dagger}_{\sigma,k}c_{\sigma,k}$, as the observable to be evaluated.
We associate $H_{d}(t)$, $H_{l}$, and $H_{t}$ with $H_{S}$, $H_{E}$, and $H_{{\rm int}}$, respectively.
Moreover, we consider a steplike change in the direction of ${\bf M}$ around the $z$-axis: dividing the period ${\cal T}$ into $N$ intervals, $t_{i}\le t'\le t_{i+1}$ ($i=1,\cdots,N$) with $t_{1}=0$ and $t_{N+1}={\cal T}$, and assume the time-dependence of $\theta(t)$ and $\phi(t)$ to be
	\begin{equation}
	\theta(t)={\rm const.},\;\;\;\phi(t)
=\biggr(1+\biggr\lfloor\frac{t}{\delta t}\biggr\rfloor\biggr)\delta\phi,
	\label{eq:steplike}
	\end{equation}
where $\lfloor x\rfloor\equiv\max\{n\in\mathbb{Z}|n\leq x\}$ is the floor function, $\delta\phi\equiv2\pi/N$, and $\delta t\equiv t_{i+1}-t_{i}={\cal T}/N$.
Specifically, we fix the direction of ${\bf M}$ during each interval $t_{i}\le t'\le t_{i+1}$, and change $\phi$ discretely at each $t_{i}$ with substitution $\phi_{i}=\phi_{i-1}+\delta\phi$ and initialization $\phi_{0}=0$, where $\phi_{i}$ is the fixed angle during the $i$th interval.
Assuming that the density matrix for the total system is factorized as $W(t_{i})=\rho(t_{i})\otimes\rho^{{\rm eq}}_{{\rm l}}$ at each $t_{i}$, Eq. (\ref{eq:formula}) yields the number of transferred electron during the interval.

Denoting the sequential difference in counting outcomes of $N_{\sigma}$ from $t_{i}$ to $t_{i+1}$ as $\Delta n_{\sigma,i}$, and the counting field associated with $N_{\sigma}$ as $\lambda_{\sigma}$, the mean number of transferred electrons with spin $\sigma$ during the time interval is given by
	\begin{equation}
	\langle\Delta n_{\sigma,i}\rangle=\int^{t_{i+1}}_{t_{i}}dt'\dbra{1}\biggr[
\frac{\partial\Xi^{(\lambda_{\sigma})}_{i}}{\partial(i\lambda_{\sigma})}
\biggr]_{\lambda_{\sigma}=0}\dket{\rho^{(0)}_{i}(t')},
	\label{eq:formula-i}
	\end{equation}
where $\Xi^{(\lambda_{\sigma})}_{i}$ and $\dket{\rho^{(0)}_{i}(t')}$ are respectively the generator and the density matrix of the quantum dot in the interval.
We also introduce its time derivative
	\begin{equation}
	J_{\sigma,i}(t)\equiv\frac{d\langle\Delta n_{\sigma,i}\rangle}{dt}=\dbra{1}\biggr[
\frac{\partial\Xi^{(\lambda_{\sigma})}_{i}}{\partial(i\lambda_{\sigma})}
\biggr]_{\lambda_{\sigma}=0}\dket{\rho^{(0)}_{i}(t)},
	\label{eq:electron-flow}
	\end{equation}
for which the time integration provides the number of transferred electrons.
By summing $\langle\Delta n_{\sigma,i}\rangle$ over one cycle of the precession, we have the total number of electrons with spin $\sigma$ transferred during the cycle,
	\begin{equation}
	\langle\Delta n_{\sigma}\rangle=\sum_{i=1}^{N}\langle\Delta n_{\sigma,i}\rangle.
	\label{eq:electron_number}
	\end{equation}

With this expression for the number of electrons transferred, we define the spin current as
	\begin{equation}
	I_{\uparrow}\equiv\frac{\langle\Delta n_{\uparrow}\rangle-
\langle\Delta n_{\downarrow}\rangle}{{\cal T}}.
	\label{eq:spin-current}
	\end{equation}
Here we have subtracted $\langle\Delta n_{\downarrow}\rangle$ from $\langle\Delta n_{\uparrow}\rangle$ in defining the spin-up current because the net number of spin-down electrons transferred from the dot to the lead reduces the net number of spin-up electrons transferred.

\section{Adiabatic limit}

We next consider spin pumping in the adiabatic limit subject to condition (\ref{eq:condition2}) following the procedure by Sinitsyn and Nemenman in Ref. \cite{sinitsyn}.
We establish that the spin current Eq. (\ref{eq:spin-current}) vanishes in the adiabatic limit.

In Ref. \cite{sinitsyn}, the adiabatic limit was assessed by dividing the cycle of modulation into interval of duration $\delta t(\equiv{\cal T}/N)$ and assuming that the system quickly approaches its steady state in each interval.
Following a similar procedure, we divide the cycle of precession into durations $\delta t$, which correspond to the steplike changes in ${\bf M}(t)$ introduced in Sec. III.B.
We evaluate the density matrix of the quantum dot during interval $t_{i}\leq t\leq t_{i+1}$ by using the completeness relation for the right- and left-eigenstates of $\Xi^{(0)}_{i}$. Assuming that the system quickly approaches its steady state, as in Ref.  \cite{sinitsyn}, we only need the term corresponding to the steady state in the spectral decomposition.
Evaluating the density matrix up to first-order in $\delta t$, we find that the first order term in $\delta t$ vanishes in our model, implying that in the adiabatic limit the density matrix becomes $\dket{\rho^{(0)}_{0}(t)}\approx\dket{u^{(0)}_{0}(t_{i})}$, where $\dket{u^{(0)}_{0}(t_{i})}$ is the steady state satisfying $\Xi^{(0)}_{i}\dket{u^{(0)}_{0}(t_{i})}=0$ (see Appendix \ref{app:adiabatic} for details of the approximation).
For steady state $\dket{u^{(0)}_{0}(t_{i})}$, we also find that there is no electron transfer between dot and lead (see Eq. (\ref{eq:d5})).
Therefore, we have
	\begin{equation}
	\langle\Delta n_{\sigma,i}\rangle\approx\int^{t_{i+1}}_{t_{i}}dt'\dbra{1}\biggr[
\frac{\partial\Xi^{(\lambda_{\sigma})}_{i}}{\partial(i\lambda_{\sigma})}
\biggr]_{\lambda_{\sigma}=0}\dket{u^{(0)}_{0}(t_{i})}=0,
	\label{eq:adiabatic_limit}
	\end{equation}
indicating that there is no net electron transfer in the interval and hence no spin current is generated in the adiabatic limit. We therefore need to analyze next nonadiabatic effects on spin pumping by the precession of magnetization.

\section{numerical evaluation of nonadiabatic spin pumping}

We present and discuss results of our numerical evaluation of spin pumping beyond the adiabatic approximation.
To describe the dot-lead coupling, we use the Ohmic spectral density with an exponential cutoff $v(\omega)\equiv\sum_{k}v_{k}^{2}\delta(\omega-\omega_{k})=\lambda\omega\exp[-\omega/\omega_{c}]$, where $\lambda$ is the coupling strength and $\omega_{c}$ the cutoff frequency.
In numerical evaluations, we set $\lambda=0.05$ and $\omega_{c}=4\omega_{\rm{u}}$ with a unit angular frequency defined as $\omega_{\rm{u}}\equiv2M/\hbar$, which corresponds to the angular frequency of the Rabi oscillation between states $\ket{0,1}$ and $\ket{1,0}$.

Introducing a unit energy $\epsilon_{\rm{u}}\equiv\hbar\omega_{\rm{u}}=2M$ and a unit time $t_{\rm u}\equiv2\pi/\omega_{\rm{u}}$, we normalize angular frequency, energy, inverse temperature, and time, and introduce the notation ${\bar\omega}\equiv\omega/\omega_{\rm{u}}$, ${\bar\epsilon}\equiv\epsilon/\epsilon_{\rm{u}}$, ${\bar\beta}\equiv\beta/\epsilon_{\rm{u}}$, and ${\bar t}\equiv t/t_{\rm{u}}$, respectively.

\subsection{Static magnetization}

Let us first analyze the electron and spin dynamics under a static magnetization setting the polar angle of the magnetization to $\theta=3\pi/4$ and its azimuthal angle to $\phi=0$.
As an initial condition, we set the dot in the Gibbs state
	\begin{equation}
	\rho(0)=e^{-\beta_{\rm{d}}H_{\rm{d}}}/Z,\;\;\;
Z={\rm{Tr}}_{\rm{d}}[e^{-\beta_{\rm{d}}H_{\rm{d}}}],
	\label{eq:initial}
	\end{equation}
where $\beta_{\rm{d}}$ is the effective inverse temperature of the dot.
In Fig. 1, we show the time evolution setting both the initial effective temperature of dot and the lead temperature to 0; that is, $\beta_{\rm{d}}^{-1}=\beta_{\rm{l}}^{-1}=0$.
This initial condition implies that the dot is empty at ${\bar t}=0$ because the lowest energy eigenstate of the dot is $\ket{0,0}$ (Sec. II).
We also set the chemical potential $\mu$ of the lead and its inverse temperature $\beta_{\rm{l}}$ to satisfy the conditions
\begin{equation}
\epsilon_{\rm{d}}-M<\mu<\epsilon_{\rm{d}}+M\;\;\;{\rm and}\;\;\;\beta_{\rm{l}}^{-1}\lesssim2M.
\label{paracondition}
\end{equation}
We chose parameter settings ${\bar\epsilon}_{\rm d}=10$ for the unpolarized energy of the dot and ${\bar\beta}_{\rm{l}}^{-1}=0$ the temperature of the lead, which satisfy conditions (\ref{paracondition}), and introduced ${\bar\mu}_{10}$ for the chemical potential of the lead.

\begin{figure}[t]\label{fig:static}
\begin{center}
\includegraphics[width=1\linewidth]{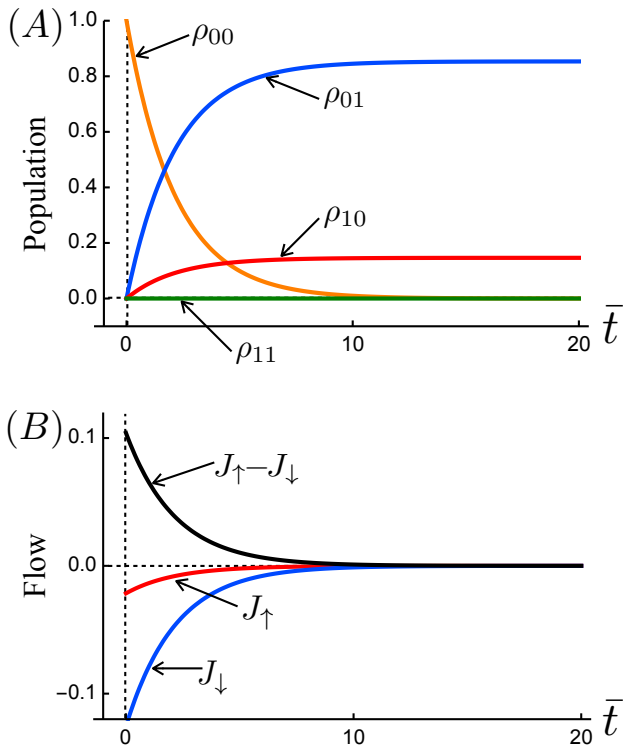}
\end{center}
\vspace{-10pt}
\caption{(A) Time evolution of the populations and (B) time-dependence of the flows are plotted as functions of the normalized time ${\bar t}$.}
\end{figure}
In Fig. 1(A), we plot the time evolution of the populations $\rho_{jj'}(t)\equiv\bra{j,j'}\rho(t)\ket{j,j'}$ ($\rho_{00}$: empty state, $\rho_{10}$: half-filled with $\uparrow$, $\rho_{01}$: half-filled with $\downarrow$, and $\rho_{11}$: completely filled).
Clearly, the electrons are flowing into the dot during the initial stage, but subsequently, within a certain time-scale (relaxation time), the populations approach their steady state values, where only the half-filled states $\rho_{10}$ and $\rho_{01}$ are populated.
The steady state corresponds to the state where a single electron with spin parallel to the magnetization ${\bf M}$ is in the quantum dot (see ${\tilde\rho}^{{\rm st}}_{10}$ in (\ref{eq:stationary-M}) in Appendix \ref{app:stationary}).
The time evolution is physically reasonable because, for the initial distribution (\ref{eq:initial}) with the condition (\ref{paracondition}), a lead electron at the Fermi level may move to an energy level of the dot located below the Fermi level, which is the state with a single electron with spin parallel to ${\bf M}$ (see Eq. (\ref{eq:hamiltonian-rotate}) in Appendix \ref{app:rotating}), due to the dot--lead interaction $H_{{\rm t}}$ to reduce the total energy of the composite system.

In Fig. 1(B), we also present the time evolution of the time derivative of the transferred electron numbers $J_{\uparrow}$ (red line), $J_{\downarrow}$ (blue line), and their difference $J_{\uparrow}-J_{\downarrow}$ (black line).
They indicate the flow of electrons from dot to lead at each moment, and in particular that more spin-$\downarrow$ electrons flow into the dot than spin-$\uparrow$ electrons.
These flows are consistent with $\rho_{01}>\rho_{10}$ from the time evolution of the populations [Fig. 1(A)].
Because $J_{\uparrow}>J_{\downarrow}$, their difference takes positive values $J_{\uparrow}-J_{\downarrow}>0$ that decay with time.
This indicates that a positive spin current is generated in the lead because the time integration of $J_{\uparrow}-J_{\downarrow}$ provides the amount of spin current generated (see Eq. (\ref{eq:spin-current})).
Note also that there is no electron current when the system is in the steady state.
This absence is analytically proved in Appendix E.

We note that the condition (\ref{paracondition}) is essential for spin pumping.
This is because, if the condition is not satisfied, either electrons are not transferring to the dot (as $\epsilon_{d}-M>\mu$) or spin-$\uparrow$ and spin-$\downarrow$ electrons of equal amounts flow onto the dot (for $\mu>\epsilon_{d}+M$).
In both cases, we have $J_{\uparrow}-J_{\downarrow}=0$, and hence there is no spin current generation.

\subsection{Precessing magnetization}

\subsubsection{Constant frequency}

Consider now a quantum dot with a precessing magnetization.
Different from the above, we suppose that the dot is initially in steady state (\ref{eq:stationary-z}) with setting $\theta=3\pi/4$ and $\phi=0$ as the starting point of a precessing magnetization instead of the Gibbs state (\ref{eq:initial}) to exclude any transient spin transfer in the initial stage of the precession.

\begin{figure}[t]\label{fig:pumping2}
\begin{center}
\includegraphics[width=1\linewidth]{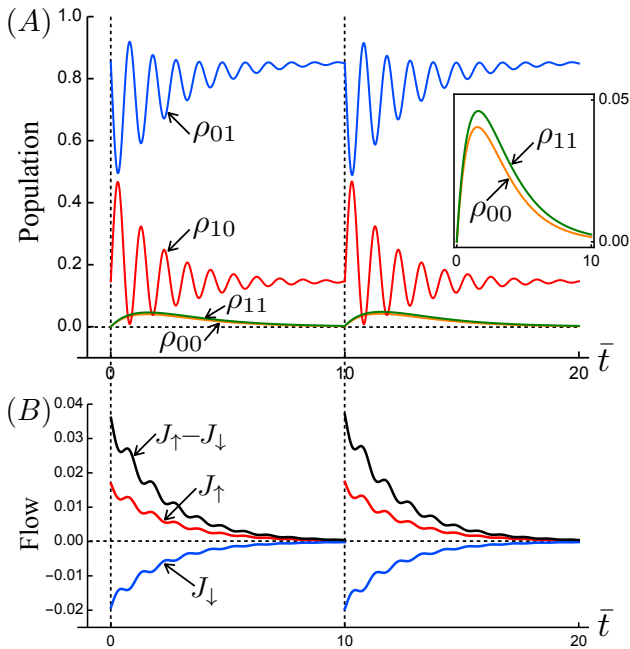}
\end{center}
\vspace{-10pt}
\caption{Time evolution of populations $J_{\sigma}$ and $J_{\uparrow}-J_{\downarrow}$ for the first two intervals are plotted. The number of divisions of the period is $N=5$.}
\end{figure}
In Fig. 2(A) and (B), we show the time evolution of populations, $J_{\sigma}$ and $J_{\uparrow}-J_{\downarrow}$ for the first two intervals.
In the calculation, the number of divisions is set to $N=5$ and the time interval to $\delta{\bar t}=10$.
Therefore the change in angle at each subsequent $t_{i}$ is $\delta\phi=2\pi/5$, that is, $\phi_{i}=\phi_{i-1}+2\pi/5$ with $\phi_0=0$.

In Fig. 2(A), we find that initially the populations deviate from their steady state values by changing $\phi$ at $t_{i}$, but then they approach steady state (\ref{eq:stationary-z}) for each $\phi_{i}$ with the populations remaining unchanged from their initial value because the steady-state populations (\ref{eq:stationary-z}) do not depend on $\phi$.
In the figure, the time evolution of the components $\rho_{01}$ and $\rho_{10}$ (blue and red lines) exhibit oscillations caused by transitions between states $\ket{0,1}$ and $\ket{1,0}$ in consequence of the applied magnetization.
Its period is given by the Rabi period between the two states $T_{R}\equiv\hbar/2M=t_{u}$.

The other two components $\rho_{00}$ and $\rho_{11}$ also show transient behavior after changing $\phi$.
We present an amplification of their time evolution [inset of Fig. 2(A)] where we find that $\rho_{00}$ and $\rho_{11}$ do not show Rabi oscillation. This is understandable because the two states $\ket{0,0}$ and $\ket{1,1}$ do not contribute to the time evolution under the magnetization [see Eq. (\ref{eq:matrix-dotH})].
While their values are small compared to $\rho_{01}$ or $\rho_{10}$, they definitely contribute to the electron transfer between dot and lead.
This is because the electron dynamics always involves a transfer via $\rho_{00}$ (or $\rho_{11}$), which we can see from the generator for the second order of the dot--lead interaction (\ref{eq:generator}) (see also (\ref{eq:matrix0}) in Appendix B).

In Fig. 2(B), the colored lines represent $J_{\sigma}$, and the black line represents their difference $J_{\uparrow}-J_{\downarrow}$ show that spin-$\uparrow$ electrons (red line) and with spin $\downarrow$ electrons (blue line) are moving in opposite directions; the former move from dot to lead whereas the latter move from lead to dot.
These trends show that the quantities $J_{\uparrow}$ and $J_{\downarrow}$ are balanced as a result of charge conservation in the lead.
Their difference (black line) is positive, $J_{\uparrow}-J_{\downarrow}>0$, indicating that this positive spin current is generated in the lead without an associated charge current.

Calculating the spin current for different values of $\theta$, we find that its spin polarization exhibits a $\theta$ dependence in that for $0<\theta<\pi/2$ the spin polarization of the spin current is anti-parallel to the $z$-axis whereas for $\pi/2<\theta<\pi$ the spin polarization is parallel to the $z$-axis.
This is because the relative relationship between the energy levels of the states $\ket{0,1}$ and $\ket{1,0}$ is interchanged; specifically, the energy level of $\ket{1,0}$ is higher than that of $\ket{0,1}$ in the former case, whereas the order of energies is the reverse in the latter case.
As the population of the lower level is larger than the other during the time evolution (see Sec. V.B.1 subsection), the spin polarization of the spin current is exchanged with respect to $\theta$.

\subsubsection{Frequency dependence}

Consider next the dependence of the spin current on the frequency of precession $\Omega=2\pi/{\cal T}$.
Here we change the period ${\cal T}=N\delta t$ by varying the time interval $\delta t$ while the number of divisions remains fixed at $N=20$.
All other parameters and initial conditions are set as before.
The dependence of the spin current on $N$ is studied in Appendix \ref{app:ndependence}.

\begin{figure}[t]
\begin{center}
\includegraphics[width=1\linewidth]{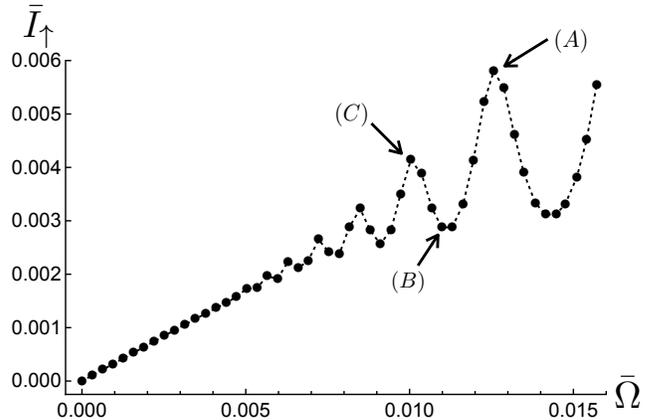}
\end{center}
\vspace{-10pt}
\caption{Frequency dependence of the generated spin current. The horizontal axis represents the normalized frequency of the magnetization ${\bar\Omega}=\Omega/\omega_{\rm{u}}$ and the vertical axis represents the normalized spin current ${\bar I}_{\uparrow}=I_{\uparrow}/\omega_{\rm{u}}$.}
\end{figure}
In Fig. 3, we plot the dependence of the generated spin current $I_{\uparrow}$ against the normalized as ${\bar\Omega}\equiv\Omega/\omega_{\rm{u}}$.
We find that the $\Omega$-dependence of the spin current features two regimes: a low-frequency regime (up to ${\bar\Omega}\sim0.005$), where $I_{\uparrow}$ depends linearly on $\Omega$, and a high-frequency regime, where $I_{\uparrow}$ oscillates depending on $\Omega$.
In the following, we explain the origins of the characteristics of the $\Omega$-dependence in the two regimes from the electron dynamics.

\begin{figure*}[t]
\begin{center}
\includegraphics[width=1\linewidth]{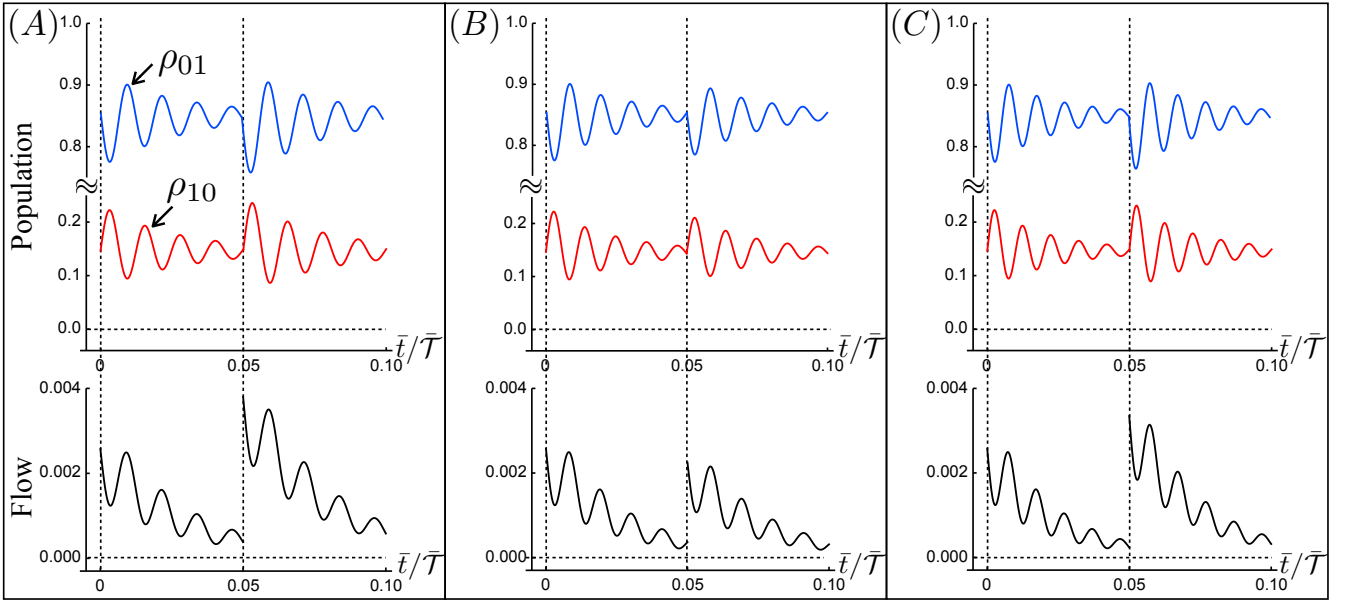}
\end{center}
\vspace{-10pt}
\caption{Three examples of the electron and spin dynamics in the high-frequency regime. The frequencies are (A) ${\bar\Omega}=0.0125$, (B) ${\bar\Omega}=0.0110$, and (C) ${\bar\Omega}=0.0100$. In each figure, the first five intervals are plotted. The horizontal axis has been normalized by the magnetization period.}
\end{figure*}
Regarding the oscillation in the high-frequency regime (Fig. 4), we plotted the time evolution of $\rho_{01}$, $\rho_{10}$ and $J_{\uparrow}-J_{\downarrow}$ at three different values of $\Omega$ as (A) ${\bar\Omega}=0.0125$, (B) ${\bar\Omega}=0.0110$, and (C) ${\bar\Omega}=0.0100$ where the values are chosen to satisfy (A) $\delta t=4 T_{R}$, (B)$\delta t=4.5 T_{R}$ and (C)$\delta t=5 T_{R}$ for the Rabi period $T_{R}=2 \pi/\omega_{u}$.
We find that the spin current takes maximum values for (A) and (C) and a minimum value for (B), (see Fig. 3).
In this regime, the time interval $\delta t$ is comparable with the relaxation time of the system, and hence $\phi$ changes during the relaxation of the Rabi oscillation.

By comparing Fig. 4(A) and (C) with (B), we find that the electron flow is enhanced by changing $\phi$ when $\delta t$ is an integer multiple of the Rabi period $T_{R}$, and is suppressed when $\delta t$ is a half-integer multiple of $T_{R}$.
This arises because the $\phi$ dependence of the off-diagonal components of $H_{\rm{d}}$, Eq. (\ref{eq:matrix-dotH}).
Since the components representing the transition from $\ket{0,1}$ to $\ket{1,0}$ and for the reverse direction contain exponentials of $\phi$ with different signs, the change of $\phi$ enhances one of these transition, and suppresses the remainder.
In the present case, the counter-clockwise precession enhances the former transition as one can see from the figures; $\rho_{10}$ increases and $\rho_{01}$ decreases just after the change in $\phi$.
This suggests a means to enhance spin pumping by changing $\phi$ in synchrony with the Rabi oscillation as for (A) and (C).

Regarding the linear dependence of $\Omega$ in the low-frequency regime, the time interval $\delta t$ is sufficiently larger than the relaxation time. Hence the populations have already reached their steady state values when $\phi$ is changed (see Fig. 2).
In this case, the net amount of spin generated during the $i$th interval
	\begin{equation}
	\langle\Delta n_{\uparrow,i}\rangle-\langle\Delta n_{\downarrow,i}\rangle
=\int^{t_{i+1}}_{t_{i}}dt\bigr[J_{\uparrow,i}(t)-J_{\downarrow,i}(t)\Bigr],
	\end{equation}
is a constant independent of the upper bound of the time integration $t_{i+1}$ because $J_{\uparrow,i}(t)-J_{\downarrow,i}(t)$ has reached $0$ at a certain time $t$ with $t<t_{i+1}$.
As the numerator in Eq. (\ref{eq:spin-current}) is constant, the spin current depends linearly on $\Omega$,
	\begin{equation}
	I_{\uparrow}=\frac{\langle\Delta n_{\uparrow}\rangle-\langle\Delta n_{\downarrow}\rangle}{2\pi}\Omega.
	\end{equation}

\section{Discussion and Concluding remarks}

We have investigated spin pumping from the view point of the adiabaticity condition for relaxation time Eq. (\ref{eq:condition2}). By formulating spin pumping using full counting statistics with the quantum master equation approach, we studied the frequency dependence of the spin current and the electron dynamics underlying it. The main results are summarized as follows:
(i) the spin current vanishes in the adiabatic limit with condition (\ref{eq:condition2}), which means that spin pumping is entirely a nonadiabatic effect in the sense of the adiabaticity condition,
(ii) the nonadiabatic spin current oscillates depending on the frequency in the high-frequency regime, reflecting the competition between the transient Rabi oscillation of the electron in the quantum dot and its relaxation time, and
(iii) the spin current depends linearly on the frequency in the low-frequency regime.

The result (i) apparently contradicts that from conventional studies, which derive adiabatic expressions for spin pumping \cite{tserkovnyak1,tserkovnyak2,wang,chen}.
This stems from the difference between the adiabaticity conditions (\ref{eq:condition1}) and (\ref{eq:condition2}).
Unlike the former condition, which solely requires the frequency to be small irrespective of the relaxation rate, the latter condition strictly requires that the relevant system is always in its steady state.
Therefore, the inference is that an adiabatic effect under the former condition includes a nonadiabatic effect in the sense of the latter condition.

As a finite relaxation time is essential to obtain result (ii), the oscillation of the spin current has not been raised in conventional studies.
The result suggests a means to enhance spin pumping by setting the frequency of precession in synchrony with the Rabi frequency of the dot electron.

Whereas a linear dependence of spin pumping on the frequency has been obtained in previous studies \cite{tserkovnyak1,tserkovnyak2,wang,chen}, it cannot be simply compared with result (iii).
This is because conventional studies have focused on contributions from spin flips between half-filled states, $\rho_{01}$ and $\rho_{10}$.
To establish a clear relationship between these studies, we need to investigate further corrections to the conventional linear formula originating from contributions from the empty and completely filled states, $\rho_{00}$ and $\rho_{11}$.
This remains a topic for future investigation.

We note that the reliability of the result for the high-frequency regime is restricted up to $\Omega\sim\tau_{R}^{-1}$, because in the evaluation we have used the Markovian approximation in obtaining Fig. 3.
To study the regime beyond this restriction, it is necessary to evaluate the counting statistics without the Markovian approximation.
Treatment of the non-Markovian effect with the full counting statistics has been studied in Ref. \cite{guarnieli}.
From this treatment, we are able to study a time scale that is sufficiently shorter than the relaxation time, which has been analyzed in conventional studies on spin pumping.
There, we expect that a clear relationship can be established between the two adiabaticity conditions (\ref{eq:condition1}) and (\ref{eq:condition2}).
We also note that our conclusion, the absence of spin pumping in the adiabatic limit, is owing to the conservation of the steady state population of the quantum dot under the precession.
The condition may not be satisfied in a more complex quantum system such as a magnetically anisotropic quantum dot \cite{SJCheng}, thus a further study on the role of nonadiabaticity in such a system is an interesting issue.
These issues remain open problems for future study.

\section*{Acknowledgements}
The authors thank Y. Tokura and S. Nakajima for valuable discussions.
This work was supported by a Grant-in-Aid for Exploratory Research (No.16K13853).

\appendix

\section{Rotating frame}\label{app:rotating}

In Sec. II, we introduced a representation of the quantum-dot Hamiltonian taking as the quantization axis for spin the $z$-axis of the lab frame.
We now introduce an another representation of this Hamiltonian introducing a basis that rotates with the magnetization ${\bf M}$---i.e., the rotating frame.
We introduce creation and annihilation operators ${\tilde d}_{s}^{\dagger}$ and ${\tilde d}_{s}$, where $s=\uparrow_{M}$ and $\downarrow_{M}$ represents the spin polarization of the electron parallel and anti-parallel, respectively, to ${\bf M}$ by
	\begin{equation}
	\left(
	\begin{array}{c}
	{\tilde d}_{\uparrow_{M}}\\
	{\tilde d}_{\downarrow_{M}}\\
	\end{array}
	\right)
	=\left(
	\begin{array}{cc}
	\cos\frac{\theta(t)}{2} & e^{-i\phi(t)}\sin\frac{\theta(t)}{2} \\
	e^{+i\phi(t)}\sin\frac{\theta(t)}{2} & -\cos\frac{\theta(t)}{2} \\
	\end{array}
	\right)
	\left(
	\begin{array}{c}
	d_{\uparrow}\\
	d_{\downarrow}\\
	\end{array}
	\right),
	\end{equation}
and their Hermitian conjugates ${\tilde d}_{\uparrow_{M}}^{\dagger}$ and ${\tilde d}_{\downarrow_{M}}^{\dagger}$.
Using these operators, the quantum dot Hamiltonian is expressible in diagonal form,
	\begin{equation}
	{\tilde H}_{{\rm d}}=\sum_{s=\uparrow_{M},\downarrow_{M}}\epsilon_{s}{\tilde d}_{s}^{\dagger}{\tilde d}_{s},
	\end{equation}
where the tilde over the Hamiltonian signifies the rotating-frame representation, $\epsilon_{\uparrow_{M}}=\epsilon_{{\rm d}}-M$ and $\epsilon_{\downarrow_{M}}=\epsilon_{{\rm d}}+M$. In the rotating frame, we use the eigenstates of $\sum_{s}{\tilde d}_{s}^{\dagger}{\tilde d}_{s}$, denoted by $\ket{{\tilde j}_{\uparrow_{M}},{\tilde j}_{\downarrow_{M}}}$ with ${\tilde j}_{s}=0$ or $1$, as a basis, in which ${\tilde H}_{{\rm d}}$ has matrix representation
	\begin{equation}
	{\tilde H_{{\rm d}}}=\left(\begin{array}{cccc}
	 0 & 0 & 0 & 0 \\
	 0 & \epsilon_{{\rm d}}+M & 0 & 0\\
	 0 & 0 & \epsilon_{{\rm d}}-M & 0\\
	 0 & 0 & 0 & 2\epsilon_{d}\\
	\end{array}
	\right).
	\label{eq:hamiltonian-rotate}
	\end{equation}

We introduce the unitary transformation $u(t)$ from $H_{{\rm d}}(t)$ to ${\tilde H}_{{\rm d}}$,
	\begin{equation}
	{\tilde H_{{\rm d}}}\equiv u(t)H_{{\rm d}}u^{\dagger}(t),
	\end{equation}
for which the matrix representation is
	\begin{equation}
	u(t)=\left(\begin{array}{cccc}
	 1 & 0 & 0 & 0 \\
	 0 & -\cos\frac{\theta}{2} & e^{+i\phi(t)}\sin\frac{\theta}{2} & 0\\
	 0 & e^{-i\phi(t)}\sin\frac{\theta}{2} & \cos\frac{\theta}{2} & 0\\
	 0 & 0 & 0 & 1\\
	\end{array}
	\right).
	\end{equation}

In the rotating frame, the interaction Hamiltonian is represented as
	\begin{equation}
	{\tilde H}_{{\rm t}}=\hbar\sum_{s}\sum_{\sigma}\sum_{k}
(w_{s,\sigma,k}{\tilde d}_{s}^{\dagger}c_{\sigma,k}
+w_{s,\sigma,k}^{*}c_{\sigma,k}^{\dagger}{\tilde d}_{s}),
	\label{eq:int_rotate}
	\end{equation}
where the coefficients $w_{s,\sigma,k}$ are given by
	\begin{eqnarray}
	\begin{split}
	&w_{\uparrow_{M},\uparrow,k}\equiv v_{k}\cos\frac{\theta}{2},\qquad
w_{\downarrow_{M},\uparrow,k}\equiv v_{k}e^{+i\phi}\sin\frac{\theta}{2},\\
	&w_{\uparrow_{M},\downarrow,k}\equiv v_{k}e^{-i\phi}\sin\frac{\theta}{2},\;w_{\downarrow_{M},\downarrow,k}\equiv -v_{k}\cos\frac{\theta}{2},
	\end{split}
	\label{eq:coefficients}
	\end{eqnarray}
and $w_{s,\sigma,k}^{*}$ is the complex conjugate of $w_{s,\sigma,k}$.

\section{A supermatrix representation of the generator}\label{app:supermatrix}

We present next a supermatrix representation of the generator, expressing it in the rotating frame instead of the lab frame.
This is because the former representation is simpler and more useful than the latter.
The representation in the lab frame can be obtained by applying a unitary transformation in the supermatrix space as discussed later in this appendix.

\begin{widetext}
In the rotating frame, the generator in the $i$th interval is defined by
	\begin{equation}
	{\tilde\xi}^{({\boldsymbol\lambda})}_{i}{\tilde\rho}\equiv\frac{1}{i\hbar}
[{\tilde H}_{{\rm{d}}},{\tilde\rho}]-\frac{1}{\hbar^{2}}\int^{\infty}_{0}
d\tau{\rm Tr}_{{\rm l}}[{\tilde H}_{{\rm t}},[{\tilde H}_{{\rm t}}(-\tau),
{\tilde\rho}\otimes\rho^{{\rm eq}}_{{\rm l}}]_{{\boldsymbol\lambda}}
]_{{\boldsymbol\lambda}}.
	\label{eq:generator-r}
	\end{equation}
where ${\boldsymbol\lambda}$ is the set of counting fields ${\boldsymbol\lambda}\equiv(\lambda_{\uparrow},\lambda_{\downarrow})$, which reduces to ${\tilde\xi}^{(\lambda_{\sigma})}_{i}$ by setting one of two counting fields equal to zero.
For the two counting fields, we define a modified operator $A^{({\boldsymbol\lambda})}\equiv e^{i(\lambda_{\uparrow}N_{\uparrow}+\lambda_{\downarrow}N_{\downarrow})}Ae^{-i(\lambda_{\uparrow}N_{\uparrow}+\lambda_{\downarrow}N_{\downarrow})}$.
The modified interaction Hamiltonian and its interaction picture are given by
\begin{equation}
{\tilde H}_{{\rm t}}^{({\boldsymbol\lambda})}=\hbar\sum_{s}\sum_{\sigma}\sum_{k}
(w_{s,\sigma,k}{\tilde d}_{s}^{\dagger}c_{\sigma,k}e^{+i\lambda_{\sigma}}
+w_{s,\sigma,k}^{*}c_{\sigma,k}^{\dagger}{\tilde d}_{s}e^{-i\lambda_{\sigma}}),
\end{equation}
\begin{equation}
{\tilde H}^{({\boldsymbol\lambda})}_{{\rm t}}(t)=\hbar\sum_{s}\sum_{\sigma}\sum_{k}
(w_{s,\sigma,k}{\tilde d}_{s}^{\dagger}c_{\sigma,k}
e^{+i\lambda_{\sigma}}e^{+i(\omega_{s}
-\omega_{k})t}+w_{s,\sigma,k}^{*}c_{\sigma,k}^{\dagger}{\tilde d}_{s}
e^{-i\lambda_{\sigma}}e^{-i(\omega_{s}-\omega_{k})t}),
\end{equation}
respectively. By inserting these expressions into Eq. (\ref{eq:generator-r}), we obtain
	\begin{eqnarray}
	\begin{split}
	{\tilde\xi}^{({\boldsymbol\lambda})}{\tilde\rho}
=&-i\sum_{s}\omega_{s}({\tilde d}_{s}^{\dagger}{\tilde d}_{s}{\tilde\rho}
-{\tilde\rho}{\tilde d}_{s}^{\dagger}{\tilde d}_{s})\\
	&-\sum_{s,s'}\Bigr[\Phi^{-(0)}_{s,s'}(t)
\tilde{d}_{s}^{\dagger}\tilde{d}_{s'}{\tilde\rho}
+\Phi^{+(0)*}_{s,s'}(t)\tilde{d}_{s}\tilde{d}_{s'}^{\dagger}{\tilde\rho}
+\Phi^{-(0)*}_{s,s'}(t){\tilde\rho}\tilde{d}_{s'}^{\dagger}\tilde{d}_{s}
+\Phi^{+(0)}_{s,s'}(t){\tilde\rho}\tilde{d}_{s'}\tilde{d}_{s}^{\dagger}\\
	&\hspace{4.4ex}-\Phi^{+({\boldsymbol\lambda})}_{s,s'}(t)\tilde{d}_{s}^{\dagger}{\tilde\rho}\tilde{d}_{s'}-\Phi^{-({\boldsymbol\lambda})*}_{s,s'}(t)\tilde{d}_{s}{\tilde\rho}\tilde{d}_{s'}^{\dagger}-\Phi^{+(-{\boldsymbol\lambda})*}_{s,s'}(t)\tilde{d}_{s'}^{\dagger}{\tilde\rho}\tilde{d}_{s}-\Phi^{-(-{\boldsymbol\lambda})}_{s,s'}(t)\tilde{d}_{s'}{\tilde\rho}\tilde{d}_{s}^{\dagger}\Bigr],
	\end{split}
	\label{eq:xi_rotate}
	\end{eqnarray}
where
	\begin{equation}
	\Phi^{\pm({\boldsymbol\lambda})}_{s,s'}(t)\equiv\sum_{\sigma}\sum_{k}\int^{t}_{0}d\tau w_{s,\sigma,k}w^{*}_{s',\sigma,k}e^{i(\omega_{s'}-\omega_{k})\tau}f^{\pm}(\epsilon_{k})e^{-i\lambda_{\sigma}},
	\end{equation}
	\begin{equation}
	f^{+}(\epsilon_{k})\equiv{\rm Tr}_{{\rm l}}[c_{\sigma,k}^{\dagger}c_{\sigma,k}\rho^{{\rm eq}}_{{\rm l}}]=\frac{1}{1+e^{\beta_{{\rm l}}(\epsilon_{k}-\mu)}},
	\end{equation}
and
	\begin{equation}
	f^{-}(\epsilon_{k})\equiv{\rm Tr}_{{\rm l}}[c_{\sigma,k}c_{\sigma,k}^{\dagger}\rho^{{\rm eq}}_{{\rm l}}]=1-f^{+}(\epsilon_{k}).
	\end{equation}
Here, the Fermi distribution of the lead electron $f^{+}(\epsilon_{k})$ is independent of $\sigma$ because the energy levels of the lead electrons with spin-up and spin-down are degenerate.

By collecting the matrix elements of the reduced density matrix ${\tilde\rho}^{({\boldsymbol\lambda})}(t)$ into the form of a vector, we obtain a supermatrix representation of the generator.
As the Hilbert space of the quantum dot has dimension 4, the density matrix has 16 elements.
Thus the density matrix is represented by a 16 dimensional vector, and the generator is represented by a $16\times16$ supermatrix.
Nevertheless, to evaluate the first cumulant of $\Delta n_{\sigma,i}$ using Eq. (\ref{eq:formula-i}), we do not need all 16 components of the density matrix because some of these components do not contribute to the trace taken in the expression.
Indeed, we find that only 6 components ${\tilde\rho}^{({\boldsymbol\lambda})}_{00}(t)\equiv\bra{{\tilde0},{\tilde0}}{\tilde\rho}^{({\boldsymbol\lambda})}(t)\ket{{\tilde0},{\tilde0}}$, ${\tilde\rho}^{({\boldsymbol\lambda})}_{01}(t)\equiv\bra{{\tilde0},{\tilde1}}{\tilde\rho}^{({\boldsymbol\lambda})}(t)\ket{{\tilde0},{\tilde1}}$, ${\tilde\rho}^{({\boldsymbol\lambda})}_{0110}(t)\equiv\bra{{\tilde0},{\tilde1}}{\tilde\rho}^{({\boldsymbol\lambda})}(t)\ket{{\tilde1},{\tilde0}}$, ${\tilde\rho}^{({\boldsymbol\lambda})}_{1001}(t)\equiv\bra{{\tilde1},{\tilde0}}{\tilde\rho}^{({\boldsymbol\lambda})}(t)\ket{{\tilde1},{\tilde0}}$, ${\tilde\rho}^{({\boldsymbol\lambda})}_{10}(t)\equiv\bra{{\tilde1},{\tilde0}}{\tilde\rho}^{({\boldsymbol\lambda})}(t)\ket{{\tilde1},{\tilde0}}$, and ${\tilde\rho}^{({\boldsymbol\lambda})}_{11}(t)\equiv\bra{{\tilde1},{\tilde1}}{\tilde\rho}^{({\boldsymbol\lambda})}(t)\ket{{\tilde1},{\tilde1}}$ contribute to the first cumulant.
By arranging the components as $\dket{{\tilde\rho}^{({\boldsymbol\lambda})}(t)}=({\tilde\rho}^{({\boldsymbol\lambda})}_{00}(t),{\tilde\rho}^{({\boldsymbol\lambda})}_{01}(t),{\tilde\rho}^{({\boldsymbol\lambda})}_{0110}(t),{\tilde\rho}^{({\boldsymbol\lambda})}_{1001}(t),{\tilde\rho}^{({\boldsymbol\lambda})}_{10}(t),{\tilde\rho}^{({\boldsymbol\lambda})}_{11}(t))^{{\rm t}}$, where $(\cdots)^{{\rm t}}$ denotes transposition, we obtain a supermatrix representation of the generator
	\begin{equation}
	{\tilde\Xi}^{({\boldsymbol\lambda})}_{i}=\left(\begin{array}{cccccc}
	X^{+(0)}_{\uparrow_{M}}+X^{+(0)}_{\downarrow_{M}} & -X^{-(-{\boldsymbol\lambda})}_{\downarrow_{M}} & Y^{-({\boldsymbol\lambda})*} & Y^{-(-{\boldsymbol\lambda})} & -X^{-({-\boldsymbol\lambda})}_{\uparrow_{M}} & 0\\
	-X^{+({\boldsymbol\lambda})}_{\downarrow_{M}} & X^{+(0)}_{\uparrow_{M}}+X^{-(0)}_{\downarrow_{M}} & 0 & 0 & 0 & -X^{-(-{\boldsymbol\lambda})}_{\uparrow_{M}}\\
	Y^{+({\boldsymbol\lambda})} & 0 & Z & 0 & 0 & Y^{-(-{\boldsymbol\lambda})}\\
	Y^{+(-{\boldsymbol\lambda})*} & 0 & 0 & Z^{*} & 0 & Y^{-({\boldsymbol\lambda})*}\\
	-X^{+({\boldsymbol\lambda})}_{\uparrow_{M}} & 0 & 0 & 0 & X^{+(0)}_{\downarrow_{M}}+X^{-(0)}_{\uparrow_{M}} & -X^{-(-{\boldsymbol\lambda})}_{\downarrow_{M}}\\
	0 & -X^{+({\boldsymbol\lambda})}_{\uparrow_{M}} & Y^{+(-{\boldsymbol\lambda})*} & Y^{+({\boldsymbol\lambda})} & -X^{+({\boldsymbol\lambda})}_{\downarrow_{M}} & X^{-(0)}_{\uparrow_{M}}+X^{-(0)}_{\downarrow_{M}}
	\end{array}
	\right),
	\label{eq:matrix}
	\end{equation}
with
	\begin{subequations}
	\begin{equation}
	X^{\pm({\boldsymbol\lambda})}_{s}\equiv\Phi^{\pm({\boldsymbol\lambda})}_{s,s}+\Phi^{\pm(-{\boldsymbol\lambda})*}_{s,s},
	\end{equation}
	\begin{equation}
	Y^{\pm({\boldsymbol\lambda})}\equiv-\Phi^{\pm({\boldsymbol\lambda})}_{\downarrow_{M},\uparrow_{M}}-\Phi^{\pm(-{\boldsymbol\lambda})*}_{\uparrow_{M},\downarrow_{M}},
	\end{equation}
and
	\begin{equation}
	Z\equiv-i\frac{2M}{\hbar}+\Phi^{+(0)*}_{\uparrow_{M},\uparrow_{M}}+\Phi^{-(0)*}_{\uparrow_{M},\uparrow_{M}}+\Phi^{+(0)}_{\downarrow_{M},\downarrow_{M}}+\Phi^{-(0)}_{\downarrow_{M},\downarrow_{M}}.
	\end{equation}
	\end{subequations}
The generator for the density matrix without a counting field is obtained by setting ${\boldsymbol\lambda}=0$ as
	\begin{equation}
	{\tilde\Xi}^{(0)}_{i}=\left(\begin{array}{cccccc}
	X^{+(0)}_{\uparrow_{M}}+X^{+(0)}_{\downarrow_{M}} & -X^{-(0)}_{\downarrow_{M}} & 0 & 0 & -X^{-(0)}_{\uparrow_{M}} & 0\\
	-X^{+(0)}_{\downarrow_{M}} & X^{+(0)}_{\uparrow_{M}}+X^{-(0)}_{\downarrow_{M}} & 0 & 0 & 0 & -X^{-(0)}_{\uparrow_{M}}\\
	0 & 0 & Z & 0 & 0 & 0\\
	0 & 0 & 0 & Z^{*} & 0 & 0\\
	-X^{+(0)}_{\uparrow_{M}} & 0 & 0 & 0 & X^{+(0)}_{\downarrow_{M}}+X^{-(0)}_{\uparrow_{M}} & -X^{-(0)}_{\downarrow_{M}}\\
	0 & -X^{+(0)}_{\uparrow_{M}} & 0 & 0 & -X^{+(0)}_{\downarrow_{M}} & X^{-(0)}_{\uparrow_{M}}+X^{-(0)}_{\downarrow_{M}}
	\end{array}
	\right).
	\label{eq:matrix0}
	\end{equation}

The reduced density matrix in the lab frame is obtained by applying the unitary transformation to ${\tilde\rho}^{({\boldsymbol\lambda})}(t)$, $\rho^{({\boldsymbol\lambda})}(t)
=u^{\dagger}(t){\tilde\rho}^{({\boldsymbol\lambda})}(t)u(t)$.
The unitary transformation on the Hilbert space operator can also be represented by a supermatrix.
For the 6 dimensional vector $\dket{{\tilde\rho}^{({\boldsymbol\lambda})}(t)}$, the unitary transformation supermatrix $U(t)$ connecting $\dket{{\tilde\rho}^{({\boldsymbol\lambda})}(t)}$ with the density matrix in the lab frame $\dket{\rho^{({\boldsymbol\lambda})}(t)}$, i.e., $\dket{{\tilde\rho}^{({\boldsymbol\lambda})}(t)}=U(t)\dket{\rho^{({\boldsymbol\lambda})}(t)}$, is given by
	\begin{equation}
	U(t)=\left(\begin{array}{cccccc}
	1 & 0 & 0 & 0 & 0 & 0\\
	0 & \cos^{2}\frac{\theta}{2} & \frac{1}{2}e^{-i\phi}\sin\theta & \frac{1}{2}e^{i\phi}\sin\theta & \sin^{2}\frac{\theta}{2} & 0\\
	0 & -\frac{1}{2}e^{i\phi}\sin\theta & \cos^{2}\frac{\theta}{2} & -e^{2i\phi}\sin^{2}\frac{\theta}{2} & \frac{1}{2}e^{i\phi}\sin\theta & 0\\
	0 & -\frac{1}{2}e^{-i\phi}\sin\theta & -e^{-2i\phi}\sin^{2}\frac{\theta}{2} & \cos^{2}\frac{\theta}{2} & \frac{1}{2}e^{-i\phi}\sin\theta & 0\\
	0 & \sin^{2}\frac{\theta}{2} & -\frac{1}{2}e^{-i\phi}\sin\theta & -\frac{1}{2}e^{i\phi}\sin\theta & \cos^{2}\frac{\theta}{2} & 0\\
	0 & 0 & 0 & 0 & 0 & 1
	\end{array}
	\right).
	\label{eq:unitary}
	\end{equation}
We can obtain a supermatrix representation of the generator in lab frame by applying the unitary supermatix to Eq. (\ref{eq:matrix}), that is, $\Xi^{({\boldsymbol\lambda})}_{i}=U^{\dagger}(t){\tilde\Xi}^{({\boldsymbol\lambda})}_{i}U(t)$.
\end{widetext}

\section{steady state}\label{app:stationary}

We now derive a steady-state solution of the master equation (\ref{eq:QME}) for $\lambda_{\sigma}=0$.
We first derive an expression for the steady-state solution in the rotating frame and then obtain it in the lab frame by applying the unitary transformation Eq. (\ref{eq:unitary}).
To derive the steady-state solution, we use the graphical method discussed in Refs. \cite{caplan82,haken}, following the latter in particular.

Let us denote the steady state in the $i$th interval as $\dket{{\tilde u}^{(0)}_{0}(t_{i})}\equiv({\tilde\rho}^{\rm{st}}_{00},{\tilde\rho}^{\rm{st}}_{01},{\tilde\rho}^{\rm{st}}_{0110},{\tilde\rho}^{\rm{st}}_{1001},{\tilde\rho}^{\rm{st}}_{10},{\tilde\rho}^{\rm{st}}_{11})^{\rm{t}}$.
It satisfies ${\tilde\Xi}^{(0)}_{i}\dket{{\tilde u}^{(0)}_{0}(t_{i})}=0$.
From the expression (\ref{eq:matrix0}), we immediately obtain ${\tilde\rho}^{{\rm st}}_{0110}=0$ and ${\tilde\rho}^{{\rm st}}_{1001}=0$.
The rest of the components of the density matrix satisfy the homogeneous equations
	\begin{widetext}
	\begin{equation}
	\left(\begin{array}{cccc}
	 X^{+(0)}_{\uparrow_{M}}+X^{+(0)}_{\downarrow_{M}} & -X^{-(0)}_{\downarrow_{M}} & -X^{-(0)}_{\uparrow_{M}} & 0 \\
	 -X^{+(0)}_{\downarrow_{M}} & X^{+(0)}_{\uparrow_{M}}+X^{-(0)}_{\downarrow_{M}} & 0 & -X^{-(0)}_{\uparrow_{M}} \\
	 -X^{+(0)}_{\uparrow_{M}} & 0 & X^{+(0)}_{\downarrow_{M}}+X^{-(0)}_{\uparrow_{M}} & -X^{-(0)}_{\downarrow_{M}} \\
	 0 & -X^{+(0)}_{\uparrow_{M}} & -X^{+(0)}_{\downarrow_{M}} & X^{-(0)}_{\uparrow_{M}}+X^{-(0)}_{\downarrow_{M}}
	\end{array}
	\right)
	\left(\begin{array}{c}
	{\tilde\rho}^{\rm{st}}_{00}\\
	{\tilde\rho}^{\rm{st}}_{01}\\
	{\tilde\rho}^{\rm{st}}_{10}\\
	{\tilde\rho}^{\rm{st}}_{11}
	\end{array}
	\right)=0.
	\label{eq:equations1}
	\end{equation}
	\end{widetext}

\begin{figure}[t]
\begin{center}
\includegraphics[width=1\linewidth]{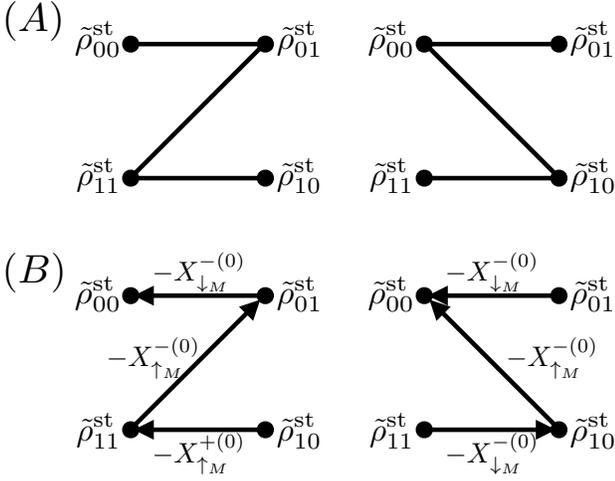}
\end{center}
\vspace{-10pt}
\caption{(A) Maximum spanning trees and (B) directed graphs associated with the node ${\tilde\rho}^{{\rm st}}_{00}$.\label{fig:graph}}
\end{figure}
The nontrivial solution of the homogeneous equations (\ref{eq:equations1}) is now obtained graphically. Consider graphs consisting of 4 nodes and 3 edges [Fig. \ref{fig:graph}(A)].
Each node represents a component of the density matrix, and each edge represents a nonzero matrix element between the components.
These graphs are called {\it maximum} {\it spanning} {\it trees} referring to graphs containing neither isolated nodes nor closed circuits.
By defining the direction of every edge with respect to a certain node, we obtain a {\it directed} {\it graph} for the node.
In Fig. \ref{fig:graph}(B), we depict the directed graphs for the node corresponding to ${\tilde\rho}^{\rm{st}}_{00}$.
The directed edge represents a transition from the node at the tail of the arrow to the node at the head of the arrow.
For each directed edge, we associate a matrix element representing the transition as a weight as presented in the figure.
By multiplying all weights in the graph, we obtain a value of the graph.
For the graph in Fig. \ref{fig:graph}(B), the values are $-X^{-(0)}_{\downarrow_{M}}X^{-(0)}_{\uparrow_{M}}X^{+(0)}_{\uparrow_{M}}$ (left) and $-X^{-(0)}_{\downarrow_{M}}X^{-(0)}_{\uparrow_{M}}X^{-(0)}_{\downarrow_{M}}$.
The sum of the values $A_{00}=-X^{-(0)}_{\downarrow_{M}}X^{-(0)}_{\uparrow_{M}}X^{+(0)}_{\uparrow_{M}}-X^{-(0)}_{\downarrow_{M}}X^{-(0)}_{\uparrow_{M}}X^{-(0)}_{\downarrow_{M}}$ is the value for the node ${\tilde\rho}^{\rm{st}}_{00}$.
Similarly, we obtain values for the other three nodes: $A_{01}=-X^{+(0)}_{\downarrow_{M}}X^{-(0)}_{\uparrow_{M}}X^{+(0)}_{\downarrow_{M}}-X^{+(0)}_{\downarrow_{M}}X^{-(0)}_{\uparrow_{M}}X^{-(0)}_{\downarrow_{M}}$, $A_{10}=-X^{-(0)}_{\downarrow_{M}}X^{+(0)}_{\uparrow_{M}}X^{-(0)}_{\downarrow_{M}}-X^{-(0)}_{\downarrow_{M}}X^{+(0)}_{\uparrow_{M}}X^{-(0)}_{\downarrow_{M}}$, and $A_{11}=-X^{+(0)}_{\downarrow_{M}}X^{+(0)}_{\uparrow_{M}}X^{+(0)}_{\downarrow_{M}}-X^{+(0)}_{\downarrow_{M}}X^{+(0)}_{\uparrow_{M}}X^{-(0)}_{\downarrow_{M}}$.
Using the values $A_{nm}$ ($n,m=0$ or $1$), the solution of Eq. (\ref{eq:equations1}) is given as
	\begin{equation}
	{\tilde\rho}^{\rm{st}}_{nm}=\frac{A_{nm}}{A_{00}+A_{01}+A_{10}+A_{11}}.
	\end{equation}
Using Eq. (B9a) with Eqs. (B5)--(C2), we obtain the steady-state solution of the master equation
	\begin{equation}
	\dket{{\tilde u}^{(0)}_{0}(t_{i})}=\left(\begin{array}{c}
	{\tilde\rho}^{\rm{st}}_{00}\\
	{\tilde\rho}^{\rm{st}}_{01}\\
	{\tilde\rho}^{\rm{st}}_{0110}\\
	{\tilde\rho}^{\rm{st}}_{1001}\\
	{\tilde\rho}^{\rm{st}}_{10}\\
	{\tilde\rho}^{\rm{st}}_{11}
	\end{array}
	\right)=\left(\begin{array}{c}
	f^{-}(\epsilon_{\uparrow_{M}})f^{-}(\epsilon_{\downarrow_{M}})\\
	f^{-}(\epsilon_{\uparrow_{M}})f^{+}(\epsilon_{\downarrow_{M}})\\
	0\\
	0\\
	f^{+}(\epsilon_{\uparrow_{M}})f^{-}(\epsilon_{\downarrow_{M}})\\
	f^{+}(\epsilon_{\uparrow_{M}})f^{+}(\epsilon_{\downarrow_{M}})
	\end{array}
	\right).
	\label{eq:stationary-M}
	\end{equation}
By applying the unitary transformation (\ref{eq:unitary}), we also obtain the steady-state solution in the lab frame $\dket{u^{(0)}_{0}(t_{i})}=(\rho^{\rm{st}}_{00},\rho^{\rm{st}}_{01},\rho^{\rm{st}}_{0110},\rho^{\rm{st}}_{1001},\rho^{\rm{st}}_{10},\rho^{\rm{st}}_{11})^{\rm{t}}$ as
	\begin{widetext}
	\begin{equation}
	\dket{u^{(0)}_{0}(t_{i})}=\left(\begin{array}{c}
	f^{-}(\epsilon_{\uparrow_{M}})f^{-}(\epsilon_{\downarrow_{M}})\\
	\cos^{2}\frac{\theta}{2}f^{+}(\epsilon_{\uparrow_{M}})f^{-}(\epsilon_{\downarrow_{M}})+\sin^{2}\frac{\theta}{2}f^{-}(\epsilon_{\uparrow_{M}})f^{+}(\epsilon_{\downarrow_{M}})\\
	e^{+i\phi(t_{i})}\cos\frac{\theta}{2}\sin\frac{\theta}{2}(f^{+}(\epsilon_{\uparrow_{M}})f^{-}(\epsilon_{\downarrow_{M}})-f^{+}(\epsilon_{\downarrow_{M}})f^{-}(\epsilon_{\uparrow_{M}}))\\
	e^{-i\phi(t_{i})}\cos\frac{\theta}{2}\sin\frac{\theta}{2}(f^{+}(\epsilon_{\uparrow_{M}})f^{-}(\epsilon_{\downarrow_{M}})-f^{+}(\epsilon_{\downarrow_{M}})f^{-}(\epsilon_{\uparrow_{M}}))\\
	\sin^{2}\frac{\theta}{2}f^{+}(\epsilon_{\uparrow_{M}})f^{-}(\epsilon_{\downarrow_{M}})+\cos^{2}\frac{\theta}{2}f^{-}(\epsilon_{\uparrow_{M}})f^{+}(\epsilon_{\downarrow_{M}})\\
	f^{+}(\epsilon_{\uparrow_{M}})f^{+}(\epsilon_{\downarrow_{M}})
	\end{array}
	\right).
	\label{eq:stationary-z}
	\end{equation}
	\end{widetext}
From these expressions, the steady-state populations in the lab frame $\rho^{\rm{st}}_{00}$, $\rho^{\rm{st}}_{01}$, $\rho^{\rm{st}}_{10}$, and $\rho^{\rm{st}}_{11}$ do not depend on $\phi$.
Thus the steady-state populations remain unchanged by changing $\phi$.

\section{Derivation of Eq. (\ref{eq:adiabatic_limit})}\label{app:adiabatic}

We now derive the expression Eq. (\ref{eq:adiabatic_limit}) in the adiabatic limit following Ref. \cite{sinitsyn}.
For the steplike precession of ${\bf M}(t)$, the density matrix of the quantum dot during $t_{i}\leq t\leq t_{i+1}$ is expressed by
	\begin{equation}
	\dket{\rho^{(0)}_{i}(t)}=e^{\Xi^{(0)}_{i}(t-t_{i})}\prod^{i-1}_{j=1}
e^{\Xi^{(0)}_{j}\delta t}\dket{\rho^{(0)}_{1}(0)},
	\label{eq:d1}
	\end{equation}
where $\dket{\rho^{(0)}_{1}(0)}$ is the initial condition in the first interval.
Similar to the procedure in Sec. V.B, we suppose that the initial condition is a steady state with $\phi=\phi_{1}-\delta\phi$, denoting it by $\dket{\rho^{(0)}_{1}(0)}=\dket{u^{(0)}_{0}(t_{0})}$ with $t_{0}=t_{1}-\delta t$.
By assuming that the system quickly approaches its steady state in each interval, we only need the term corresponding to the steady state in the spectral decomposition of each interval in Eq. (\ref{eq:d1}).
Hence we obtain
	\begin{equation}
	\dket{\rho^{(0)}_{i}(t)}\approx\dket{u^{(0)}_{0}(t_{i})}\prod^{i}_{j=1}\dbraket{v^{(0)}_{0}(t_{j})}{u^{(0)}_{0}(t_{j-1})},
	\label{eq:d2}
	\end{equation}
where $\dbra{v^{(0)}_{0}(t_{j})}$ is the left zero eigenstate of $\Xi^{(0)}_{j}$ in the $j$th interval satisfying $\dbra{v^{(0)}_{0}(t_{j})}\Xi^{(0)}_{j}=0$.
The left zero eigenstate is given by the trace operator as $\dbra{v^{(0)}_{0}(t_{j})}=\dbra{1}$ because of the trace invariance $\dbraket{1}{\rho^{(0)}_{j}(t)}=1$ (see comment below Eq. (17)).
Evaluating $\dbraket{v^{(0)}_{0}(t_{j})}{u^{(0)}_{0}(t_{j-1})}
=\dbraket{v^{(0)}_{0}(t_{j})}{u^{(0)}_{0}(t_{j}-\delta t)}$ in Eq. (\ref{eq:d2}) up to first order in $\delta t$, we obtain
	\begin{equation}
	\dket{\rho^{(0)}_{i}(t)}\approx\dket{u^{(0)}_{0}(t_{i})}\prod^{i}_{j=1}
\Bigr(1-\dbra{v^{(0)}_{0}(t_{j})}\frac{\partial}{\partial t}\dket{u^{(0)}_{0}(t)}
\Bigr|_{t=t_{j}}\delta t\Bigr).
	\label{eq:d3}
	\end{equation}
Using the expression for the steady state Eq. (\ref{eq:stationary-z}) together with $\dbra{v^{(0)}_{0}(t_{j})}=\dbra{1}$, we find $\dbra{v^{(0)}_{0}(t_{j})}\frac{\partial}{\partial t}\dket{u^{(0)}_{0}(t)}|_{t=t_{j}}=0$.
Hence we obtain the density matrix in the adiabatic limit as $\dket{\rho^{(0)}_{i}(t)}\approx\dket{u^{(0)}_{i}(t_{i})}$.
Regarding the result of Eq. (\ref{eq:formula-i}), we obtain an expression for the number of transferred electrons during the $i$th interval in the adiabatic limit,
	\begin{equation}
	\langle\Delta n_{\sigma,i}\rangle\approx\int^{t_{i+1}}_{t_{i}}dt'\dbra{1}\biggr[
\frac{\partial\Xi^{({\boldsymbol\lambda})}_{i}}{\partial(i\lambda_{\sigma})}
\biggr]_{\lambda_{\sigma}=0}\dket{u^{(0)}_{0}(t_{i})}.
	\label{eq:d4}
	\end{equation}
As the trace operation is unchanged by a unitary transformation, we obtain the same expression in the rotating frame.
Using expression (\ref{eq:matrix}) with Eq.  (\ref{eq:coefficients}), we have
	\begin{eqnarray}
	\begin{split}
	&\dbra{1}\biggr[\frac{\partial\Xi^{({\boldsymbol\lambda})}_{i}}{\partial(i\lambda_{\sigma})}\biggr]_{\lambda_{\sigma}=0}\dket{u^{(0)}_{0}(t_{i})}\\
	&=\dbra{1}\biggr[\frac{\partial{\tilde\Xi}^{({\boldsymbol\lambda})}_{i}}{\partial(i\lambda_{\sigma})}\biggr]_{\lambda_{\sigma}=0}\dket{{\tilde u}^{(0)}_{0}(t_{i})}\\
	&=2\pi\biggr\{v(\omega_{\uparrow_{M}})[f^{+}(\epsilon_{\uparrow_{M}})({\tilde\rho}^{{\rm st}}_{00}+{\tilde\rho}^{{\rm st}}_{01})\\
	&\hspace{0.8cm}-f^{-}(\epsilon_{\uparrow_{M}})({\tilde\rho}^{{\rm st}}_{10}+{\tilde\rho}^{{\rm st}}_{11})]\cos^{2}\frac{\theta}{2}\\
	&\hspace{0.8cm}+v(\omega_{\downarrow_{M}})[f^{+}(\epsilon_{\downarrow_{M}})({\tilde\rho}^{{\rm st}}_{00}+{\tilde\rho}^{{\rm st}}_{10})\\
	&\hspace{0.8cm}-f^{-}(\epsilon_{\downarrow_{M}})({\tilde\rho}^{{\rm st}}_{01}+{\tilde\rho}^{{\rm st}}_{11})]\sin^{2}\frac{\theta}{2}\biggr\},
	\end{split}
	\end{eqnarray}
where $v(\omega)=\sum_{k}v_{k}^{2}\delta(\omega-\omega_{k})$ with $\omega_{k}\equiv\epsilon_{k}/\hbar$ is the spectral density describing the dot--lead coupling, and $\omega_{s}\equiv\epsilon_{s}/\hbar$.
As the steady state (\ref{eq:stationary-M}) satisfies the relations 
$\rho^{{\rm st}}_{00}+\rho^{{\rm st}}_{01}=f^{-}(\epsilon_{\uparrow_{M}})$, $\rho^{{\rm st}}_{10}+\rho^{{\rm st}}_{11}=f^{+}(\epsilon_{\uparrow_{M}})$, $\rho^{{\rm st}}_{00}+\rho^{{\rm st}}_{10}=f^{-}(\epsilon_{\downarrow_{M}})$, and $\rho^{{\rm st}}_{01}+\rho^{{\rm st}}_{11}=f^{+}(\epsilon_{\downarrow_{M}})$,
we find
	\begin{equation}
	\dbra{1}\biggr[
\frac{\partial\Xi^{({\boldsymbol\lambda})}_{i}}{\partial(i\lambda_{\sigma})}
\biggr]_{\lambda_{\sigma}=0}\dket{u^{(0)}_{0}(t_{i})}=0,
	\label{eq:d5}
	\end{equation}
and hence we obtain the final result Eq. (\ref{eq:adiabatic_limit}).
The result Eq. (\ref{eq:d5}) also provides an analytical proof of the absence of electron flow, $J_{\sigma}=0$, in the steady state.

\section{Dependence of the spin current on the division number $N$}\label{app:ndependence}

\begin{figure}[t]
\begin{center}
\includegraphics[width=1\linewidth]{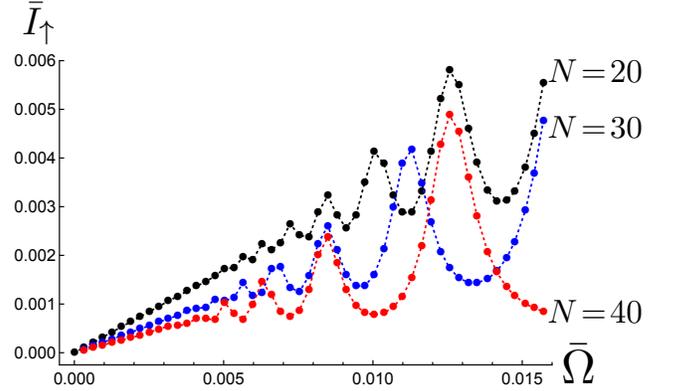}
\end{center}
\vspace{-10pt}
\caption{Spin current for three different $N$. The black, blue, and red lines correspond to $N=20$, $30$, and $40$, respectively.\label{fig:ndep}}
\end{figure}
Here we present the dependence of the spin current on the number of divisions $N$.
In Fig. \ref{fig:ndep}, we show the $\Omega$-dependence of the spin current for three numbers of divisions: $N=20, 30, 40$.
The other parameters are the same as set for calculations in Sec. V.
For each $N$, the frequency ${\bar\Omega}=2\pi/(N\delta{\bar t})$ is changed by varying the time interval $\delta{\bar t}$.
The figure shows that the amount of spin current generated decreases as the number of divisions $N$ increases.

For fixed frequency $\Omega$, the time interval $\delta t$ becomes smaller as $N$ increases.
In the limit $N\gg1$, the time interval is small as $\delta t\ll1$ whenever the adiabatic approximation Eq. (\ref{eq:d4}) is valid.
Therefore, the spin current approaches zero for large $N$.


\begin{thebibliography}{99}

\bibitem{wolf}
S. A. Wolf, D. D. Awschalom, R. A. Buhrman, J. M. Daughton, S. von Moln\'ar, M. L. Roukes, A. Y. Chtchelkanova, and D. M. Treger, Science {\bf 294}, 1488 (2001).

\bibitem{maekawa}
S. Maekawa, H. Adachi, K. Uchida, J. Ieda, and E. Saitoh, J. Soc. Phys. Jpn. {\bf 82}, 102002 (2013).

\bibitem{tserkovnyak1}
Y. Tserkovnyak, A. Brataas, and G. E. W. Bauer, Phys. Rev. Lett. {\bf 88}, 117601 (2002).

\bibitem{tserkovnyak2}
Y. Tserkovnyak, A. Brataas, and G. E. W. Bauer, Phys. Rev. B {\bf 66}, 224403 (2002).

\bibitem{wang}
B. Wang, J. Wang, and H. Guo, Phys. Rev. B {\bf 67}, 092408 (2003).

\bibitem{pzhang}
P. Zhang, Q. K. Xue, and X. C. Xie, Phys. Rev. Lett. {\bf 91}, 196602 (2003).

\bibitem{fransson}
J. Fransson and M. Galperin, Phys. Rev. B {\bf 81}, 075311 (2010).

\bibitem{winkler}
N. Winkler, M. Governale, and J. K\"onig, Phys. Rev. B {\bf 87}, 155428 (2013).

\bibitem{rojek1}
S. Rojek, M. Governale, and J. K\"onig, Phys. Status Solidi B {\bf 251}, 1912 (2013).

\bibitem{jahn}
B. O. Jahn, H. Ottosson, M. Galperin, and J. Fransson, ACS Nano {\bf 7}, 1064 (2013).

\bibitem{chen}
K. Chen and Z. Zhang, Phys. Rev. Lett. {\bf 114}, 126602 (2015).

\bibitem{tatara}
G. Tatara, Phys. Rev. B {\bf 94}, 224412 (2016).

\bibitem{mucciolo}
E. R. Mucciolo, C. Chamon, and C. M. Marcus, Phys. Rev. Lett. {\bf 89}, 146802 (2002).

\bibitem{cota}
E. Cota, R. Aguado, and G. Platero, Phys. Rev. Lett. {\bf 94}, 107202 (2005).

\bibitem{rojek2}
S. Rojek, J. K\"onig, and A. Shnirman, Phys. Rev. B {\bf 87}, 075305 (2013).

\bibitem{riwar}
R. Riwar and J. Splettstoesser, Phys. Rev. B {\bf 82}, 205308 (2010).

\bibitem{braun}
M. Braun and G. Murkard, Phys. Rev. Lett. {\bf 101}, 036802 (2008).

\bibitem{nakajima}
S. Nakajima, M. Taguchi, T. Kubo, and Y. Tokura, Phys. Rev. B {\bf 92}, 195420 (2015).

\bibitem{splettstoesser}
J. Splettstoesser, M. Governale, and J. K\"onig, Phys. Rev. B {\bf 77}, 195320 (2008).

\bibitem{brouwer}
P. W. Brouwer, Phys. Rev. B {\bf 58}, R10135 (1998).

\bibitem{moskalets}
M. Moskalets and M. B\"uttiker, Phys. Rev. B {\bf 66}, 205320 (2002).

\bibitem{moskalets_text}
M. Moskalets, {\it Scattering matrix approach to non-stationary quantum transport}, (Imperial College Press, London, 2011).

\bibitem{sinitsyn}
N. A. Sinitsyn and I. Nemenman, Europhys. Lett. {\bf 77}, 58001 (2007).

\bibitem{hanggi}
J. Ren, P. H\"anggi, and B. Li, Phys. Rev. Lett. 104, 170601 (2010).

\bibitem{yuge}
T. Yuge, T. Sagawa, A. Sugita, and H. Hayakawa, Phys. Rev. B {\bf 86}, 235308 (2012).

\bibitem{uchiyama}
C. Uchiyama, Phys. Rev. E {\bf 89}, 052108 (2014).

\bibitem{levitov}
L. S. Levitov and G. B. Lesovik, JETP Lett. {\bf 68}, 230 (1993).

\bibitem{utsumi}
Y. Utsumi, Phys. Rev. B {\bf 75}, 035333 (2007).

\bibitem{esposito}
M. Esposito, U. Harbola, and S. Mukamel, Rev. Mod. Phys. {\bf 81}, 1665 (2009).

\bibitem{kubo}
R. Kubo, J. Math. Phys. {\bf 4}, 174 (1963).

\bibitem{shibataetal}
F. Shibata, Y. Takahashi, and N. Hashitsume, J. Stat. Phys. {\bf 17}, 171 (1977).

\bibitem{hashitsume}
N. Hashitsume, F. Shibata, and M. Shingu, J. Stat. Phys. {\bf 17}, 155 (1977).

\bibitem{uchiyama99}
C. Uchiyama and F. Shibata, Phys. Rev. E {\bf 60}, 2636 (1999).

\bibitem{guarnieli}
G. Guarnieri, C. Uchiyama, and B. Vacchini, Phys. Rev. A {\bf 93}, 012118 (2016).

\bibitem{SJCheng}
S. Cheng, Phys. Rev. B {\bf 79}, 245301 (2009).

\bibitem{caplan82}
S. R. Caplan and D. Zeilberger, Adv. Appl. Math. {\bf 3}, 377(1982).

\bibitem{haken}
H. Haken, {\it Synergetics: An Introduction: Nonequilibrium Phase Transitions and Self-Organization in Physics, Chemistry, and Biology}, (Springer-Verlag, New York, 1983).

\end{thebibliography}
\end{document}